\documentclass[5p]{elsarticle}
\makeatletter
\def\ps@pprintTitle{%
 \let\@oddhead\@empty
 \let\@evenhead\@empty
 \def\@oddfoot{}%
 \let\@evenfoot\@oddfoot}
\makeatother

\usepackage{csquotes}
\usepackage{graphicx,color}
\usepackage{float}
\usepackage[caption=false]{subfig}
\usepackage{amsmath}
\usepackage{amssymb}
\usepackage{multirow}
\usepackage{dsfont}
\usepackage{adjustbox}
\usepackage{pdflscape}
\usepackage{wasysym}

\begin{document}

\title{Analytical expressions for thermophysical properties of solid and liquid beryllium relevant for fusion applications}
\author{P. Tolias}
\address{Space and Plasma Physics - KTH Royal Institute of Technology, Teknikringen 31, 10044 Stockholm, Sweden }
\begin{abstract}
\noindent The status of the literature is reviewed for thermophysical properties of pure polycrystalline solid and liquid beryllium which constitute input for the modelling of intense plasma-surface interaction phenomena that are important for fusion applications (thermal analysis, vapor shielding, melt motion, arcing, dust generation, dust transport). Reliable experimental data are analyzed for the latent heats, specific isobaric heat capacity, electrical resistivity, thermal conductivity, mass density, vapor pressure, work function, total hemispherical emissivity and absolute thermoelectric power from the room temperature up to the normal boiling point of beryllium as well as for the surface tension and the dynamic viscosity across the liquid state. Analytical expressions are recommended for the temperature dependence of these thermophysical properties, which involve high temperature extrapolations given the absence of extended liquid beryllium measurements.
\end{abstract}
\maketitle

\section{Introduction}

\noindent For next generation tokamak devices, the lifetime and power handling capabilities of the metallic plasma-facing components (PFC) constitute major design challenges\,\cite{introduct1}. ITER (currently under construction) will feature a beryllium (Be) first wall and a full tungsten (W) divertor\,\cite{introduct2,introduct3}. In the W divertor, intense steady state heat loads will elevate the surface temperatures in the vicinity of the W recrystallization range, raising concerns for edge melting and full surface melting during downward vertical displacement events (VDEs), edge-localized modes (ELMs) or loss of detachment control\,\cite{introduct3,introduct4,introduct5}. On the Be main chamber, the surface temperatures could be strongly elevated mainly during the thermal and the current quench of unmitigated VDEs or major disruptions\,\cite{introduct6,introduct7}. The resulting liquid metal layers are subject to macroscopic motion owing to the volumetric Lorentz force that is driven by replacement currents\,\cite{introduct8} but also to various instabilities that could lead to droplet ejection\,\cite{introduct9}. The thermodynamic, magneto-hydrodynamic and thermoelectric PFC response to transient events depends on numerous thermophysical material properties, whose temperature dependence needs to be accurately known a priori for simulation tools to provide reliable predictions.

In a previous work, state-of-the-art measurements of the thermophysical properties of high purity isotropic bulk tungsten were reviewed and accurate empirical expressions were proposed for their temperature dependence from the room temperature up to the W normal (atmospheric pressure) boiling point\,\cite{introductM}. The present work is focused on reviewing state-of-the-art measurements of the thermophysical properties of high purity polycrystalline bulk beryllium from the room temperature up to its normal boiling point. The thermophysical properties of interest are the latent heats of phase transitions (fusion, hcp-to-bcc polymorphic transition, vaporization), the specific isobaric heat capacity, the electrical resistivity, the thermal conductivity, the mass density, the vapor pressure, the work function, the total hemispherical emissivity and the absolute thermoelectric power (for the solid and the liquid phases) as well as the surface tension and the dynamic viscosity (only for the liquid phase). The primary objective is to identify and to critically evaluate reliable experimental datasets in order to propose accurate empirical expressions for the temperature dependence of these thermophysical quantities that will standardize their description in the multiple thermal analysis\,\cite{introcode1,introcode2}, vapor shielding\,\cite{introcode3,introcode4,introcode5}, macroscopic melt motion\,\cite{kthmemos01,kthmemos02,kthmemos03}, arcing\,\cite{introcode6,introcode7,introcode8}, dust generation\,\cite{introcode9,introcode0} and dust transport codes\,\cite{migraine01,migraine02,dusttranp1,dusttranp2,dusttranp3,dusttranp4} that are being developed by the fusion community. In the case when no reliable experimental datasets are available, the objective is to propose analytical extrapolations that do not violate general statistical mechanics principles and are consistent with well-established semi-empirical rules.

It has been possible to construct accurate analytical expressions for the thermophysical properties of solid beryllium based solely on reliable experimental data. On the other hand, narrow or wide extrapolations have to be performed for most of the thermophysical properties of liquid beryllium. Nevertheless, the attempted extrapolations are based on established empirical expressions accurate for most liquid metals and, when possible, are cross-checked with rigorous statistical mechanics constraints. The sparsity of liquid beryllium experimental data brings forth the necessity for new measurements. Fusion-specific complications due to plasma irradiation, chemical reactions, impurity alloying and surface roughness are briefly discussed.

\section{Thermophysical properties}\label{collection}

\noindent Even in the course of the most extreme off-normal tokamak events, the plasma pressures remain well below the MPa range. In addition, the $\sim100\,$GPa inverse isothermal compressibility of Be\,\cite{Thermogen1} and the $\sim0.1\,$K/MPa initial slope of the melting curve of Be\,\cite{Thermogen2} suggest that, under tokamak conditions, the thermophysical and electrochemical material properties are practically independent of the pressure and depend solely on the temperature. This removes one thermodynamic dimension from our consideration\,\cite{Thermogen3} and allows the exploitation of available systematic low pressure measurements for the extraction of reliable fits. Naturally, focus lies on high purity $>99\%$ isotropic bulk beryllium.

The evaluation of experimental values is an inherently judgmental process, but effort has been put so that the analysis remains as objective as possible. Given the large number of available data compilations, synthetic reference data and associated reviews for some thermophysical properties, the general strategy adopted in the present investigation is the following. The starting step involves the identification of the most contemporary comprehensive review of available data for each physical property, which typically contains either a recommended synthetic dataset or a recommended fitting formula. The second step involves the consideration of all experimental data or empirical expressions that were acquired after the aforementioned review. The third step involves a comparison between these independent data sources, during which any outlier data are examined for possible deviation causes. Such outlier data are also cross-checked against rigorous statistical mechanics constraints and, in case of systematic differences, are removed from consideration. It should be noted that for many properties (surface tension, dynamic viscosity, total hemispherical emissivity, absolute thermoelectric power), no reliable starting point reviews have been published in the literature and all available data had to be considered.

\subsection{Phase transition temperatures and latent heats}\label{latent}

\noindent Phase transition temperatures and latent heats are relevant for plasma-surface interaction problems that require thermal analysis and involve melting or resolidification. There is a strong consensus in the literature concerning the Be phase transition temperatures at atmospheric pressure and the associated latent heats\,\cite{Thermodyn1,Thermodyn2,Thermodyn3,Thermodyn4}. Here, we shall follow the recent recommendations of Arblaster\,\cite{Thermodyn4}. Beryllium undergoes a polymorphic phase transition from a hcp solid (alpha Be phase) to a bcc solid (beta Be phase) which takes place at $T_{\mathrm{t}}=1543\pm5\,$K and a melting phase transition from a bcc solid to a liquid which takes place at $T_{\mathrm{m}}=1560\pm5\,$K, whereas the liquid gas phase transition takes place at the normal boiling point $T_{\mathrm{b,n}}=2750\,$K. On the other hand, Be critical temperature estimates are far more uncertain and generally vary within $5400-9200\,$K\,\cite{CriticalP1} with $T_{\mathrm{c}}\simeq8080\,$K considered here\,\cite{CriticalP2}. Note the elevated pressure at the critical point $p_{\mathrm{c}}=0.05-1.22\,$GPa, which again exhibits large variations\,\cite{CriticalP1}.

The difference between the specific enthalpy of the hcp and bcc phase at the polymorphic phase transition is defined as the enthalpy of transition $\Delta{h}_{\mathrm{t}}$ and the difference between the specific enthalpy of the bcc and liquid phase at the melting point is defined as the enthalpy of fusion $\Delta{h}_{\mathrm{f}}$. The experimental determination of the individual latent heats has led to vastly different values. However, the observation that the latent heat ratio is constant around $0.85$ together with accurate measurements of the latent heat sum $\sim14.8\,$kJ/mol, suggest that the distinct latent heats can be reliably inferred\,\cite{Thermodyn1,Thermodyn2,Thermodyn3,Thermodyn4,Thermodyn5}. Here, we shall follow the recommendations $\Delta{h}_{\mathrm{t}}+\Delta{h}_{\mathrm{f}}=14.814\,$kJ/mol and $\Delta{h}_{\mathrm{t}}/\Delta{h}_{\mathrm{f}}=0.86$\,\cite{Thermodyn4}, which yield $\Delta{h}_{\mathrm{t}}=6.855\,$kJ/mol and $\Delta{h}_{\mathrm{f}}=7.959\,$kJ/mol.

In tokamak-relevant melting simulations, any attempt to resolve the very narrow beta phase temperature window of $17\pm10\,$K would drastically increase the computational cost without increasing the accuracy due to the unavoidable plasma boundary condition uncertainties. Therefore, it is preferable to ignore the polymorphic transition and combine the enthalpy of transition $\Delta{h}_{\mathrm{t}}$ with the enthalpy of fusion $\Delta{h}_{\mathrm{f}}$ in a single latent heat contribution $\Delta{h}_{\mathrm{t,f}}=14.814\,$kJ/mol that is absorbed or released exactly at the melting point. As a further consequence, all the hcp-to-bcc discontinuities in the thermophysical properties (anyways unavailable for all properties except from the specific heat capacity) can be ignored and all the hcp expressions can be extrapolated beyond the narrow bcc temperature range.

The enthalpy of vaporization $\Delta{h}_{\mathrm{v}}$ of all elements has a very weak temperature dependence\,\cite{Thermodyn6}. The same applies for Be, whose room temperature $\Delta{h}_{\mathrm{v}}$ is $324\,$kJ/mol\,\cite{Thermodyn4} and whose normal boiling temperature $\Delta{h}_{\mathrm{v}}$ is $292\,$kJ/mol \cite{Thermodyn7}. The enthalpy of vaporization can thus be considered to be temperature independent, but can also be easily parameterized based on the fact that it should be exactly zero at the critical temperature. In fact, most proposed correlations involve a weak power law with respect to $(T_{\mathrm{c}}-T)$ that includes a number of adjustable coefficients\,\cite{Thermodyn8}. We shall follow Watson's parameterization and adjust its exponent from $0.38$ to $0.28$\,\cite{Thermodyn9}, so that $\Delta{h}_{\mathrm{v}}(T)$ intersects the aforementioned known data points. Overall, we have
\begin{align*}
\Delta{h}_{\mathrm{v}}(T)=\Delta{h}_{\mathrm{v}}(T_0)\left(\frac{T_{\mathrm{c}}-T}{T_{\mathrm{c}}-T_0}\right)^{0.28}\,,
\end{align*}
with $T_0=300\,$K for the reference temperature.

It is worth comparing our selected $\Delta{h}_{\mathrm{t,f}}$ value with the prediction of Richard's rule that allows a rough estimate of $\Delta{h}_{\mathrm{t,f}}$ with knowledge of the melting temperature $T_{\mathrm{m}}$ alone\,\cite{LatentRul1,LatentRul2,LatentRul3}. Given the stability of two Be allotropes at low pressures, a modified version of Richard's rule is appropriate here, which states that the cumulative entropy of fusion and transformation is a quasi-universal constant for all metals with an approximate value $\Delta{s}_{\mathrm{t,f}}\simeq{R}$, where $R=N_{\mathrm{A}}k_{\mathrm{B}}$ is the ideal gas constant whose arithmetic value is $R=8.314\,$J/(mol$\cdot$K)\,\cite{LatentRul4}. This leads to $\Delta{h}_{\mathrm{t,f}}=T_{\mathrm{m}}\Delta{s}_{\mathrm{t,f}}$ and to the prediction $\Delta{h}_{\mathrm{t,f}}=12.970\,$kJ/mol that is very close to our recommendation.

It is also worth comparing our selected $\Delta{h}_{\mathrm{v}}(T_{\mathrm{b}})$ value with the prediction of Trouton's rule that allows a rough estimate of $\Delta{h}_{\mathrm{v}}$ with knowledge of the normal boiling temperature $T_{\mathrm{b}}$ alone\,\cite{LatentRul3}. The original version of the Trouton rule states that the entropy of vaporization is a quasi-universal constant for all metals with an approximate value $\Delta{s}_{\mathrm{v}}\simeq10.5{R}$, but it is known to provide rather crude estimates\,\cite{LatentRul5,LatentRul6,LatentRul7}. The modified Trouton's rule of Kistiakowsky has a theoretical origin, leads to more accurate estimates and reads as $\Delta{s}_{\mathrm{v}}=4.4{R}+R\ln{T_{\mathrm{b}}}$\,\cite{LatentRul8}. This leads to $\Delta{h}_{\mathrm{v}}=T_{\mathrm{b}}\Delta{s}_{\mathrm{v}}$ and to the prediction $\Delta{h}_{\mathrm{v}}=282\,$kJ/mol that is also very close to our recommendation.

\subsection{Specific isobaric heat capacity}\label{heatcapacity}

\begin{figure}[!t]
         \centering
         \includegraphics[width=3.3in]{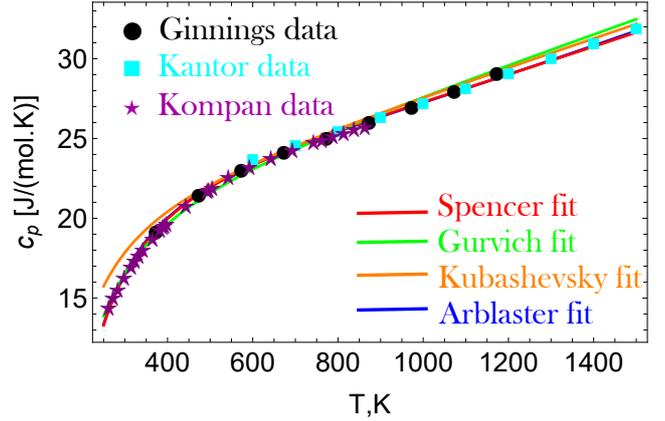}
\caption{The Be alpha phase specific isobaric heat capacity in the temperature range $300-1543\,$K according to four empirical analytical expressions and three independent measurement sets.}\label{figure_heatcapacity_comparison}
\end{figure}

\noindent The specific isobaric heat capacity is important for plasma-surface interaction problems that require thermal analysis. For the alpha phase in the temperature range $300-1543\,$K, four fitting formulas, all of the Shomate type but of different polynomial degree, are available in the literature which are essentially based on the same measurements; the nine Ginnings \emph{et al.} enthalpy data in the range $370-1170\,$K\,\cite{Thermodyn0} and the eleven Kantor \emph{et al.} enthalpy data within $600-1500\,$K\,\cite{Thermodyn5}. In the following expressions, $c_{\mathrm{p}}$ is measured in J/(mol$\cdot$K) and $T$ in Kelvin. The \emph{Spencer fit} reads as\,\cite{heatcapac1}
\begin{align*}
c_{\mathrm{p}}(T)&=21.205+5.694\times10^{-3}T+0.962\times10^{-6}\times{T}^2\\&\,\,\,\,\,\,\,-\frac{0.5874\times10^6}{T^2}\,.
\end{align*}
The \emph{Gurvich fit} reads as\,\cite{heatcapac2}
\begin{align*}
c_{\mathrm{p}}(T)&=18.736+9.293\times10^{-3}T-\frac{0.4501\times10^6}{T^2}\,.
\end{align*}
The \emph{Kubashevsky fit} reads as\,\cite{Thermodyn7}
\begin{align*}
c_{\mathrm{p}}(T)&=19.013+8.877\times10^{-3}T-\frac{0.3434\times10^6}{T^2}\,.
\end{align*}
The \emph{Arblaster fit} reads as\,\cite{Thermodyn4}
\begin{align*}
c_{\mathrm{p}}(T)&=21.539+4.9457\times10^{-3}T+1.35632\times10^{-6}\times{T}^2\\&\,\,\,\,\,\,\,-\frac{0.59408\times10^6}{T^2}\,.
\end{align*}
Recent systematic experiments were carried out by Kompan \emph{et al.} in the range of $260-870\,$K\,\cite{heatcapac3}. Their specific isobaric heat capacity measurements were only provided in a figure that was digitized with the aid of software. This additional dataset will serve as a further benchmark for the selection of the optimal fitting expression. In figure \ref{figure_heatcapacity_comparison}, the four fitting formulas have been plotted together with the three independent measurement sets. It is evident that the Gurvich fit systematically fails to capture the high temperature behavior of the specific isobaric heat capacity, that the Kubashevsky fit systematically fails to capture the low and high temperature behavior of the specific isobaric heat capacity and that the Spencer \& Arblaster fits are nearly indistinguishable. Overall, we recommend the Spencer fit owing to its smaller mean absolute relative deviations from the Kompan \emph{et al.} data that were not utilized for the construction of any of the four fitting formulas.

For the beta phase in the temperature range $1543-1560\,$K, no measurements are available. However, there is a consensus in the literature that the specific isobaric heat capacity can be considered to be constant in this narrow temperature interval with a value\,\cite{Thermodyn1,Thermodyn2,Thermodyn4}
\begin{align*}
c_{\mathrm{p}}(T)&=30.00\,\mathrm{J/(mol}\cdot\mathrm{K})\,.
\end{align*}

For the liquid phase in the temperature range $1560-2200\,$K, two linear fitting formulas are available in the literature which are both based on the sixteen enthalpy data that Kantor \emph{et al.} obtained within the temperature interval $1560-2150\,$K\,\cite{Thermodyn5}. Given the same functional form and the same data points, the Arblaster fit\,\cite{Thermodyn4} and the Chase fit\,\cite{Thermodyn2} are essentially identical. We recommend the \emph{Chase fit} that reads as\,\cite{Thermodyn2}
\begin{align*}
c_{\mathrm{p}}(T)&=25.4345+2.150\times10^{-3}T\,,
\end{align*}
where $c_{\mathrm{p}}$ is measured in J/(mol$\cdot$K) and $T$ in Kelvin. The Chase fit will be extrapolated at higher temperatures.

\begin{figure}[!t]
         \centering
         \includegraphics[width=3.3in]{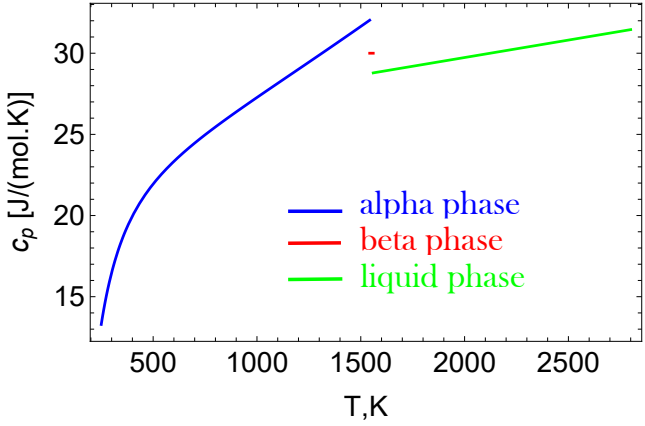}
\caption{Recommended description of the Be \emph{specific isobaric heat capacity} in the temperature range $300-2750\,$K. The $c_{\mathrm{p}}$ discontinuities at the polymorphic and liquid-solid phase transitions are visible.}\label{figure_heatcapacity_recommended}
\end{figure}

Summing\,up, our recommended description of the beryllium specific isobaric heat capacity comprises the Spencer fit for the alpha phase, a constant value for the beta phase and the Chase fit for the liquid phase. Our recommended description is plotted in Figure \ref{figure_heatcapacity_recommended}.

\subsection{Electrical resistivity}\label{electricalresistivity}

\noindent The electrical resistivity is important in plasma surface interaction problems that involve the propagation of currents inside PFCs, since they follow the least resistive path. In particular, these concern the computation of the local replacement current that is triggered by charge loss due to thermionic emission (which leads to the Lorentz force that drives macroscopic melt motion)\,\cite{kthmemos01} and the computation of the global halo current that is generated during disruptions (leading to strong electro-mechanical loads on the vessel and its components)\,\cite{resishalo1,resishalo2}.

For the solid phase of Be within the range $300-1560\,$K, a synthetic dataset is available that has been constructed from all the reliable electrical resistivity measurements\,\cite{resistivi1}. The Chi compilation consists of $20$ data points in the temperature range of interest and is recommended by a number of handbooks, see e.g. Refs.\cite{resistivi2,resistivi3}. This dataset has been fitted to a cubic polynomial that is centered around the room temperature and reads as
\begin{align*}
&\rho_{\mathrm{e}}(T)=3.71002+30.4119\times10^{-3}(T-T_0)\\&\,\,\,\,+2.7851\times10^{-6}(T-T_0)^2+3.25184\times10^{-9}(T-T_0)^3\,,
\end{align*}
with $\rho_{\mathrm{e}}$ measured in $\mu\Omega-$cm, $T$ in Kelvin and $T_0=300\,$K. The fit has a mean absolute relative error of merely $0.23\%$.

For the liquid phase in an extended temperature range, the only relevant experiments are those of Boivineau \emph{et al.} \cite{resistivi4}, who carried out measurements within $293-2062\,$K. Unfortunately, their solid phase measurements exhibit $25\%$ mean absolute relative deviations from the reliable synthetic Chi data. More important, the point-to-point deviations increase with the temperature and even reach $52\%$ close to the Be melting point, implying that these electrical resistivity measurements cannot be trusted in the Be liquid phase. Taking into account that the Boivineau \emph{et al.} mass density measurements\,\cite{resistivi4} also seem to be erroneous (see subsection \ref{massdensity}), it is possible that the errors originate from the correction of the resistivity data for thermal expansion effects, with the latter being overestimated due to the misinterpretation of hazy shadow contours as expanded wire. After excluding these experiments, only the electrical resistivity at the melting point of Be is available\,\cite{resistivi5,resistivi6}. On the basis of the small values of the resistivity temperature coefficient for most liquid metals, it is assumed that the value remains constant in the entire liquid phase,
\begin{align*}
&\rho_{\mathrm{e}}(T)=45\,\mu\Omega\,\mathrm{cm}\,.
\end{align*}

\begin{figure}[!t]
         \centering
         \includegraphics[width=3.3in]{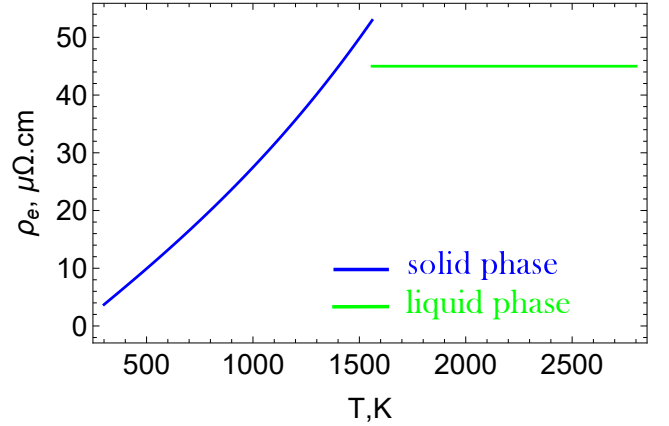}
\caption{Recommended description of the Be \emph{electrical resistivity} in the temperature range $300-2750\,$K. The $\rho_{\mathrm{e}}$ discontinuity at the liquid-solid phase transition is clearly visible.}\label{figure_electricalresistivity_recommended}
\end{figure}

Summing up, our recommended description of the beryllium electrical resistivity consists of a cubic fit for the solid phase (alpha and beta phases) and of a constant value for the liquid phase. Our recommended description is plotted in Figure \ref{figure_electricalresistivity_recommended}.

\subsection{Thermal conductivity}\label{thermalconductivity}

\noindent The thermal conductivity is relevant for all plasma-surface interaction phenomena that require thermal analysis. Dust transport constitutes the only exception, since, owing to the small dust size and the large metallic thermal diffusivity, the characteristic timescale of heat diffusion is orders of magnitude smaller than the characteristic timescale of dust motion. This implies that, at any point, spatial dust temperature gradients can be assumed be instantaneously smoothed out which amounts to ignoring heat conduction.

\begin{figure}[!b]
         \centering
         \includegraphics[width=3.3in]{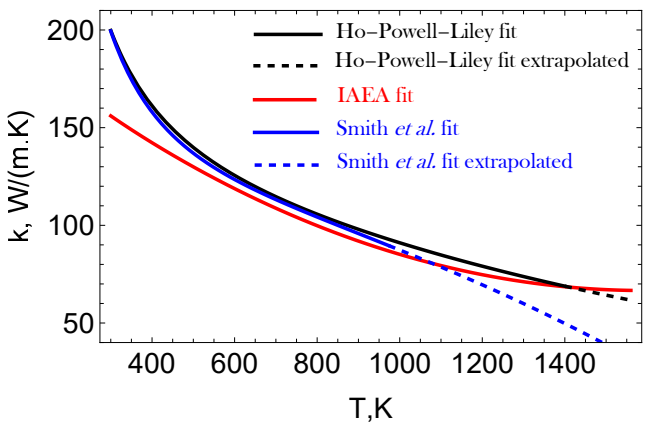}
\caption{The Be solid phase thermal conductivity in the temperature range $300-1560\,$K according to three empirical analytical expressions that originate from three independent measurement sets.}\label{figure_thermalconductivity_comparison}
\end{figure}

For the solid phase of Be within the range $300-1560\,$K, we shall analyze three thermal conductivity datasets that are available in the literature, namely; the synthetic dataset of Ho, Powell \& Liley that consists of $25$ data points from $300\,$K to $1400\,$K\,\cite{conductiv1}, the IAEA recommended dataset that has been adopted from the Soviet literature and consists of $14$ data points from $293\,$K up to $1560\,$K\,\cite{Thermodyn7}, the measurement set of Smith \emph{et al.} that consists of $11$ data points from $325\,$K up to $973\,$K, but also features two extrapolated values at $1127\,$K and $1270\,$K\,\cite{conductiv2}. Identical fits of the Shomate type have been carried out for the three datasets. In the following expressions, $k$ is measured in W/(m$\cdot$K) and $T$ in Kelvin. The \emph{Ho,\,Powell \& Liley fit} reads as
\begin{align*}
k(T)&=148.8912-76.3780\times10^{-3}T+12.0174\times10^{-6}T^2\\&\,\,\,\,\,\,\,+\frac{6.5407\times10^6}{T^2}
\end{align*}
with $0.38\%$ mean absolute relative deviations from the respective data. The \emph{IAEA fit} reads as
\begin{align*}
k(T)&=200.268-168.102\times10^{-3}T+52.846\times10^{-6}T^2\\&\,\,\,\,\,\,\,+\frac{0.12919\times10^6}{T^2}
\end{align*}
with $0.19\%$ mean absolute relative deviations from the respective data. For completeness, we note that the IAEA dataset was accompanied by a simple quadratic fit\,\cite{Thermodyn7} that has been disregarded. The \emph{Smith et al.\,fit} reads as
\begin{align*}
k(T)&=111.138+5.41524\times10^{-3}T-37.300\times10^{-6}T^2\\&\,\,\,\,\,\,\,+\frac{8.1029\times10^6}{T^2}
\end{align*}
with $0.96\%$ mean absolute relative deviation from the respective data.  In figure \ref{figure_thermalconductivity_comparison}, the three fitting formulas have been plotted together. It is evident that the IAEA fit has significant deviations from the other two fits in the entire temperature range, including the vicinity of the room temperature. It is also evident that the Ho,\,Powell \& Liley fit and the Smith \emph{et al.} fit are nearly identical up to the extrapolated values of the latter dataset. Hence, it is highly likely that the extrapolation of the Smith \emph{et al.} dataset to high temperatures is unreliable. Therefore, we recommend the use of the Ho,\,Powell \& Liley fit.

For the liquid phase in an extended temperature range, there are no thermal conductivity measurements available. An empirical closed-form expression will be constructed on the basis of the Wiedemann-Franz law\,\cite{conductiv3,conductiv4} which states that the ratio of the thermal conductivity to the electrical conductivity of metals is proportional to the temperature, $k(T)=L_0T/\rho_{\mathrm{e}}(T)$ where $L_0=2.443\times10^{-8}\,\mathrm{W}\Omega\mathrm{K}^{-2}$ is known as the nominal Lorenz number. The Wiedemann-Franz law becomes very accurate at high temperatures where heat is primarily conducted by electron-lattice interactions and at very low temperatures where heat is primarily conducted by electron-impurity interactions. For Be, the lattice contribution to the thermal conductivity could be substantial which implies a non-negligible temperature dependence of the Lorenz number $L$ and potentially significant deviations from its nominal value $L_0$\,\cite{conductiv5,conductiv6}. To determine the value of the ratio $L/L_0$ in the temperature range $300-1560\,$K, the recommended solid Be descriptions of the electrical resistivity and the thermal conductivity were employed. The Lorenz ratio attained values within $(0.90,1.15)$, which suggests that the nominal Lorenz number can be safely utilized in the liquid phase. The combination of the Wiedemann-Franz law with the recommended liquid Be description of the electrical resistivity and the recasting of the outcome as a first order expansion around the melting point led to the empirical expression
\begin{align*}
k(T)&=84.59+54.22\times10^{-3}(T-T_{\mathrm{m}})\,,
\end{align*}
with $k$ measured in W/(m$\cdot$K), $T$ in Kelvin, $T_{\mathrm{m}}=1560\,$K. An alternative closed-form expression can be constructed on the basis of Powell's empirical formula for the connection between the thermal conductivity and the electrical conductivity of solid Be\,\cite{conductiv5,conductiv6}. The empirical law is nominally valid within $323-1000\,$K and reads as\,\cite{conductiv5,conductiv6}
\begin{align*}
k(T)&=\frac{T}{\rho_{\mathrm{e}}(T)}\left(2.57-\frac{249}{T}\right)+\frac{268}{T}-0.151\,,
\end{align*}
with $k$ measured in W/(m$\cdot$K) and $\rho_{\mathrm{e}}$ measured in $\mu\Omega-$cm. The extrapolation of the Powell empirical formula in the liquid phase and the combination with the recommended liquid Be description of the electrical resistivity leads to the alternative expression, that is nearly indistinguishable from the initial expression, validating our procedure.

\begin{figure}[!b]
         \centering
         \includegraphics[width=3.3in]{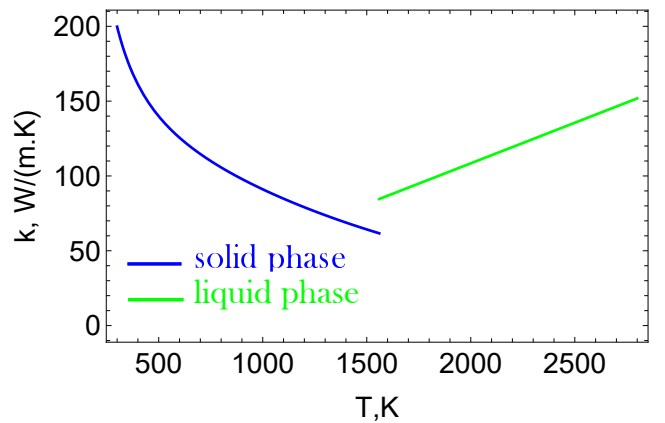}
\caption{Recommended description of the Be \emph{thermal conductivity} in the temperature range $300-2750\,$K. The $k$ discontinuity at the liquid-solid phase transition is visible.}\label{figure_thermalconductivity_recommended}
\end{figure}

Summing up, our recommended description of the beryllium thermal conductivity consists of a Shomate type fit to the Ho, Powell \& Liley dataset for the solid phase (alpha and beta phases) and a linear fit based on the Wiedemann-Franz law for the liquid phase. Our recommended description is plotted in Figure \ref{figure_thermalconductivity_recommended}.

\subsection{Mass density}\label{massdensity}

\noindent The mass density is naturally important for any plasma surface interaction problem that either requires thermal analysis or involves bulk PFC motion and dust transport. For the solid phase in the temperature range $300-1560\,$K, a synthetic dataset is available that has been constructed from all the reliable thermal expansion measurements\,\cite{massdensi1}. The Touloukian compilation consists of $10$ data points for the relative linear expansion $\Delta{L}/L_0(\%)$ within the temperature range of $400-1500\,$K. Note that Touloukian provides one cubic fit for $\Delta{L}/L_0$ around $293\,$K that is valid within the range $293-895\,$K and one cubic fit for $\Delta{L}/L_0$ around $895\,$K that is valid within the range $895-1500\,$K. However, the dependence of $\Delta{L}/L_0$ on the temperature is very smooth from the room temperature up to the melting temperature. Thus, Touloukian's piecewise fitting was judged as an unnecessary complexity. The original relative linear expansion $\Delta{L}/L_0(\%)$ dataset was first converted to a linear expansion dataset via $L/L_0=1+0.01\Delta{L}/L_0(\%)$ that was then converted to a volume expansion dataset via $V/V_0=(L/L_0)^3$ that was finally converted to a mass density dataset via $\rho_{\mathrm{m}}=\rho_{\mathrm{m0}}(V_0/V)$ with $\rho_{\mathrm{m0}}=1.850\,$g/cm${^3}$ for the Be mass density at room temperature. The latter dataset was fitted into a single cubic polynomial around $T_0=300\,$K which reads as
\begin{align*}
&\rho_{\mathrm{m}}(T)=1.850-6.8648\times10^{-5}(T-T_0)\\&\,\,\,\,\,\,\,-4.1660\times10^{-8}(T-T_0)^2+1.1354\times10^{-11}(T-T_0)^3\,,
\end{align*}
with $\rho_{\mathrm{m}}$ measured in g/cm${^3}$, $T$ in Kelvin and with $0.002\%$ mean absolute relative deviations from the respective data. A direct mass density dataset, that features $13$ data points, is available from Smith \emph{et al.} who carried out Be thermal expansion measurements in the range from $325\,$K up to $1270\,$K\,\cite{conductiv2}. Given the stated $99.4\%$ purity of their Be samples, it is evident that the authors employed an erroneous room temperature mass density of $\rho_{\mathrm{m0}}=1.820\,$g/cm${^3}$ when converting their relative linear expansion dataset to a mass density dataset. After rescaling their mass densities with the use of the correct room temperature mass density of $\rho_{\mathrm{m0}}=1.850\,$g/cm${^3}$, the corrected dataset is fitted into a single cubic polynomial around $T_0=300\,$K. The resulting fit is nearly indistinguishable from the Touloukian fit.

For the liquid phase in an extended temperature range, one relative volume expansion dataset\,\cite{resistivi4} and one mass density fitting expression\,\cite{massdensi2} are available. Similar to the electrical resistivity, the Boivineau \emph{et al.} dataset within the solid phase temperature range exhibits very large deviations from our recommended description and will, thus, not be considered in the temperature range of the liquid phase. On the other hand, the Steinberg empirical expression is recommended in a number of handbooks\,\cite{Thermodyn7,resistivi5}. It is a linear polynomial around the melting point\,that\,reads\,as
\begin{align*}
&\rho_{\mathrm{m}}(T)=1.690-0.116\times10^{-3}(T-T_{\mathrm{m}})\,,
\end{align*}
with $\rho_{\mathrm{m}}$ in g/cm${^3}$, $T$ in Kelvin and $T_{\mathrm{m}}=1560\,$K.

\begin{figure}[!t]
         \centering
         \includegraphics[width=3.3in]{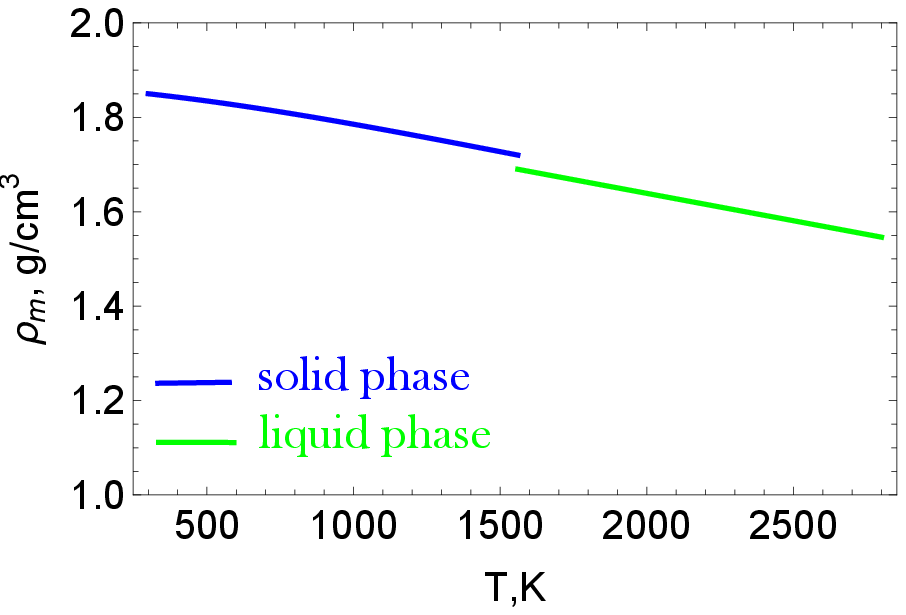}
\caption{Recommended description of the Be \emph{mass density} in the temperature range $300-2750\,$K. The $\rho_{\mathrm{m}}$ discontinuity at the liquid-solid phase transition is visible.}\label{figure_massdensity_recommended}
\end{figure}

Summing up, our recommended description of the beryllium mass density consists of a cubic polynomial around $T_0=300\,$K that is determined by fitting to the Touloukian dataset for the solid phase (alpha and beta) and a linear expansion around the melting point for the liquid phase. Our recommended description is plotted in Figure \ref{figure_massdensity_recommended}.

\subsection{Surface tension}\label{surfacetension}

\noindent Surface tension plays a key role in metallic melt dynamics, melt layer stability and even droplet break-up.  Within the shallow water approximation, the tangential stress balance boundary condition yields a thermo-capillary flow that is proportional to the temperature derivative of the surface tension\,\cite{kthmemos03} and has proven to be non-negligible for certain PFC exposure geometries\,\cite{surftensi1}. Moreover, the surface tension is the main stabilizing factor against the development of Rayleigh-Taylor or geometry-driven instabilities\,\cite{introduct9,introcode0}. Furthermore, the surface tension also constitutes the primary stabilizing factor against electrostatic and rotational droplet breakup\,\cite{surftensi2}.

For the liquid phase, only $2$ surface tension data points are available in the literature\,\cite{Thermodyn7,resistivi5}. These are $\sigma=1100$ mN/m at $1773\,$K and $\sigma=1145\,$mN/m at $1553\,$K (super-cooled state). They suffice to determine the standard linear fit around the melting temperature, which reads as
\begin{align*}
&\sigma(T)=1.143-0.20\times10^{-3}(T-T_{\mathrm{m}})\,,
\end{align*}
with $\sigma$ measured in N/m, $T$ in Kelvin and $T_{\mathrm{m}}=1560\,$K. Given the unique measurement dataset and the minimum number of data points available, it is important to devise an independent way to confirm that the above empirical expression is physical. This is possible by first extrapolating the empirical expression at high temperatures and then determining the critical point temperature from $\sigma=0$. The result is $T_{\mathrm{c}}\simeq7275\,$K, which is relatively close to the $T_{\mathrm{c}}\simeq8080\,$K estimate\,\cite{CriticalP2}.

\begin{figure}[!t]
         \centering
         \includegraphics[width=3.3in]{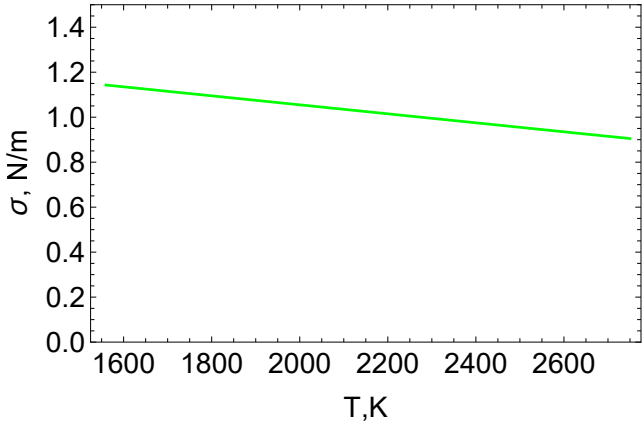}
\caption{Recommended description of the Be \emph{surface tension} in the temperature range $1560-2750\,$K.}\label{figure_surfacetension_recommended}
\end{figure}

Summing up, our recommended description of the beryllium surface tension consists of a linear expansion around the melting point that is determined by fitting to the only two data available. Our recommended description is plotted in Figure \ref{figure_surfacetension_recommended}.

\subsection{Dynamic viscosity}\label{viscosity}

\noindent Dynamic viscosity plays a key role in metallic melt dynamics and stability. Within the shallow water approximation, the viscous Navier-Stokes term yields a melt-depth dependent frictional shear force that directly counteracts the driving Lorentz force and thus leads to the emergence of a terminal melt speed\,\cite{kthmemos02,kthmemos03}. In addition, the melt viscosity is the primary stabilizing factor against the development of Kelvin-Helmholtz instabilities\,\cite{introduct9,viscosity0}.

For the liquid phase, dynamic viscosity measurements are available in a figure of Ref.\cite{viscosity1}. The dynamic viscosity data were automatically extracted after digitizing this figure with the aid of software. The Be data were fitted to a standard Arrhenius fit  $\mu(T)=\mu_0\exp{\left[\left(E_{\mathrm{a}}/RT_{\mathrm{m}}\right)\left(T_{\mathrm{m}}/T\right)\right]}$ with $E_{\mathrm{a}}$ the activation energy for viscous flow, $\mu_0$ the pre-exponential viscosity and $R=8.314\,$J/(mol$\cdot$K) the ideal gas constant\,\cite{resistivi5,viscosity2,viscosity3,viscosity4,viscosity5}. The expression reads as
\begin{align*}
&\mu(T)=0.514\times10^{-3}\exp{\left(4.635\frac{T_{\mathrm{m}}}{T}\right)}\,,
\end{align*}
where $\mu$ is measured in Pa$\cdot$s, $T$ in Kelvin and $T_{\mathrm{m}}=1560\,$K. The dynamic viscosity values seem to be too large even for a liquid metal\,\cite{resistivi5}. Thus, it is rather imperative to devise independent ways to confirm that the results are accurate.

The \emph{Fowler-Born-Green} viscosity formula emerges by combining two approximate expressions that stem from the statistical mechanics of simple liquids; the Fowler formula for the surface tension\,\cite{viscosith1,viscosith2} and the Born-Green formula for the dynamic viscosity\,\cite{viscosith3}. It reads as $\mu(T)=(16/15)\sqrt{m/(k_{\mathrm{B}}T)}\sigma(T)$, with $m$ the atomic mass\,\cite{viscosith4,viscosith5,viscosith6}, and has proved to be accurate for elemental liquid metals\,\cite{viscosity2}. The Fowler-Born-Green prediction for the Be viscosity with input from our recommended Be surface tension expression is $50$ to $15$ times smaller than the output of the empirical expression (depending on the temperature). The \emph{Andrade formula} for the melting point viscosity emerges by considering atomic momentum transfer between neighboring layers due to vibrational
displacements and using Lindemann's melting criterion\,\cite{viscosity3,viscosith7}. It reads as $\mu(T_{\mathrm{m}})=C_{\mathrm{A}}\{T_{\mathrm{m}}^{1/2}[\rho_{\mathrm{m}}(T_{\mathrm{m}})]^{2/3}/M^{1/6}\}$ with $M$ the molar mass,  $C_{\mathrm{A}}=1.8\times10^{-7}\mathrm{kg}^{1/2}\mathrm{ms}^{-1}\mathrm{K}^{-1/2}\mathrm{(mol)}^{-1/6}$ for the proportionality constant\,\cite{viscosity3,viscosith7} and has proved to be accurate for elemental liquid metals\,\cite{viscosity2,viscosity3}. The Andrade prediction for the Be melting point viscosity is nearly $25$ times smaller than the output of the empirical expression.

These theoretical tests suggest that the above empirical expression should be abandoned. An alternative empirical expression reads as\,\cite{viscosity5}
\begin{align*}
&\mu(T)=0.1\times10^{-3}\exp{\left(3.93\frac{T_{\mathrm{m}}}{T}\right)}\,,
\end{align*}
where $\mu$ is measured in Pa$\cdot$s, $T$ in Kelvin and $T_{\mathrm{m}}=1560\,$K. It exhibits a satisfactory connection with both the Fowler-Born-Green formula and the Andrade formula.

\begin{figure}[!b]
         \centering
         \includegraphics[width=3.3in]{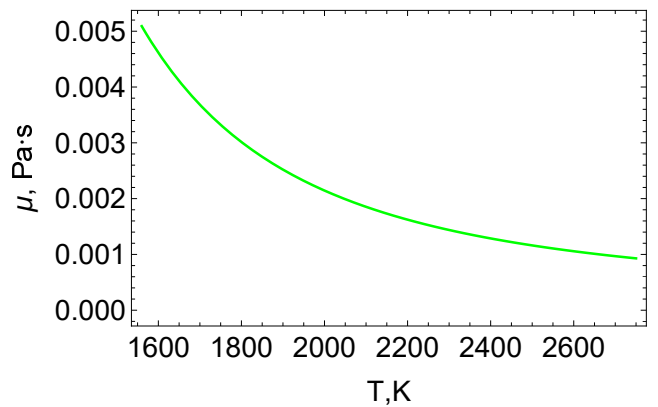}
\caption{Recommended description of the Be \emph{dynamic viscosity} in the temperature range $1560-2750\,$K.}\label{figure_dynamicviscosity_recommended}
\end{figure}

Summing up, our recommended description of the beryllium dynamic viscosity consists of an available standard Arrhenius fit. Our recommended description is plotted in Figure \ref{figure_dynamicviscosity_recommended}.

\subsection{Vapor pressure}\label{vaporpressure}

\noindent Vapor pressure is important for impurity generation from PFCs and dust as well as for the passive cooling of PFCs and dust. As described in the Knudsen formula, the evaporated mass flux is directly proportional to the vapor pressure. In addition, vaporization constitutes an omnipresent cooling channel that is directly proportional to the vapor pressure and becomes dominant beyond the melting temperature\,\cite{kthmemos03,migraine03}. Furthermore, vapor shielding constitutes an indirect cooling channel that strongly depends on the vapor pressure\,\cite{vaporpres1} and becomes dominant at temperatures somewhat lower than the normal boiling point\,\cite{vaporpres2}.

For the solid and liquid phases, a unique semi-empirical Antoine correlation will be followed for the temperature dependence of the vapor pressure. The Antoine equation for Be reads as\,\cite{vaporpres3}
\begin{align*}
p_{\mathrm{v}}(T)=10^{\displaystyle10.2089-\displaystyle\frac{13696.6102}{T-124.63}}\,,
\end{align*}
where $p_{\mathrm{v}}$ is measured in Pa and $T$ is in Kelvin. The strict validity range lies between $1097\,$K and $2757\,$K, but the expression remains very accurate when extrapolated to lower and higher temperatures. It is important to point out that Arblaster provides separate empirical correlations of the form $\ln{(p_{\mathrm{v}},\mathrm{bar})}=A+B\ln{T}+C/T+DT+ET^2$ for the alpha solid phase ($750-1525\,$K), the beta solid phase ($1543-1560\,$K) and the liquid phase ($1600-2750\,$K)\,\cite{Thermodyn4}. The Arblaster correlations are nearly indistinguishable from our preferred correlation in their entire validity range and will not be preferred being more complex. It is also important to note that the normal boiling point prediction of the preferred correlation, \emph{i.e.} the solution of $p_{\mathrm{v}}(T)=101325$, is $2757\,$K, which is highly consistent with the $T_{\mathrm{b,n}}=2750\,$K choice. Other Antoine correlations available in the literature are either valid in a narrow range or unreliable\,\cite{Thermodyn7}.

\begin{figure}[!b]
         \centering
         \includegraphics[width=3.3in]{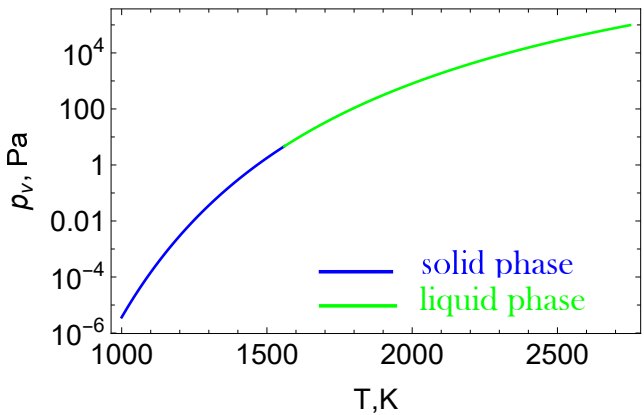}
\caption{Recommended description of the Be \emph{vapor pressure} in the temperature range $1000-2750\,$K.}\label{figure_vaporpressure_recommended}
\end{figure}

Summing up, our recommended description of the beryllium vapor pressure in the solid and liquid phase consists of a unique standard Antoine correlation. Our recommended description is plotted in Figure \ref{figure_vaporpressure_recommended}.

\subsection{Work function}\label{workfunction}

\noindent The work function plays a key role in melt layer motion and dust / droplet survivability by controlling thermionic emission, see the Richardson-Dushman law\,\cite{workfunc00}. Concerning melt dynamics, surface charge loss owing to thermionic emission triggers a bulk replacement current density that is responsible for the volumetric Lorentz force which drives macroscopic melt layer motion in tokamaks\,\cite{kthmemos01}. Concerning dust survivability, thermionic emission controls the onset of thermal instability that occurs due to the strong coupling between dust charging and heating within the orbital motion limited approach; thermionic emission shifts the dust charge towards more positive values, thus enhancing electron heating and increasing the dust surface temperature, which leads to even more enhanced thermionic emission\,\cite{dusttranp1}. Finally, thermionic emission also constitutes an important cooling channel, especially for grounded PFCs and less for floating dust/droplets\,\cite{introduct8,kthmemos03}.

Early elevated temperature measurements of work functions, based on thermionic emission in ultra-high vacuum conditions, reported $W_{\mathrm{f}}=3.6-3.9\,$eV for clean high purity bulk Be samples\,\cite{workfunc01,workfunc02,workfunc03}. Similar recommendations are found in earlier reviews\,\cite{workfunc04}. Contemporary room temperature measurements, based either on the Kelvin probe method or the photoelectric effect, reported $W_{\mathrm{f}}=4.98-5.08\,$eV for ultra-high vacuum deposited Be films\,\cite{workfunc05,workfunc06}. The latter works attributed the earlier erroneous measurements to the diffusion of oxygen impurities from the bulk volume onto the surface of the Be sample. Nowadays, there is a strong consensus in the literature that the room temperature Be work function is $W_{\mathrm{f}}=4.98\,$eV\,\cite{workfunc07,workfunc08,workfunc09}.

It should be noted that the work function depends on the temperature mainly due to thermal expansion and lattice vibration effects\,\cite{workfunc00}. In most cases, the temperature correction is small and can be expressed by a linear dependence $W_{\mathrm{f}}(T)=W_{\mathrm{f,0}}(T)+\alpha_{\mathrm{wf}}T$, where $\alpha_{\mathrm{wf}}$ lies in the range of a few tens to few hundred $\mu$eV/K and can have either negative or positive sign\,\cite{workfunc10}. Such minor corrections can have observable effects on the thermionic or thermal-field current emitted from metals. However, the deviations are of the order of other work function uncertainties (patch effects, adsorbates) and, thus, the temperature dependence of the work function is generally neglected.

\subsection{Total hemispherical emissivity}\label{emissivity}

\noindent The quantity of interest is the total hemispherical emissivity $\epsilon_{\mathrm{T}}$, \emph{i.e.} the emissivity averaged over all wavelengths and hemispherical directions. The total hemispherical emissivity is the material dependent coefficient of the Stefan-Boltzmann law of cooling owing to thermal radiation\,\cite{emissivit1}. Its importance lies in the fact that thermal radiation is a significant PFC / dust cooling channel below melting\,\cite{kthmemos03}. There is a large volume of experimental data that is available for the total normal emissivity $\epsilon_{0^{\circ}}$ of many materials, \emph{i.e.} the emissivity at the normal direction averaged over all wavelengths, but the same does not apply for $\epsilon_{\mathrm{T}}$\,\cite{emissivit2,emissivit3}. This also holds true for Be\,\cite{emissivit4,emissivit5}. It should be noted that these two emissivities can be equated only when the angular distribution of the emitted radiation follows Lambert's cosine law, which is very rarely the case. Otherwise, an assumed equivalence might lead to erroneous estimates of radiative heat losses.

For the Be solid phase in the temperature range $300-1560\,$K, a synthetic dataset is available that has been constructed from reliable total hemispherical emissivity measurements\,\cite{Thermodyn7,emissivit6}. The dataset consists of $14$ data points and was fitted with a cubic polynomial around the room temperature. The empirical expression reads as
\begin{align*}
&\epsilon_{\mathrm{T}}(T)=0.043865+0.05728\times10^{-3}(T-T_0)\\&\,-0.218399\times10^{-6}(T-T_0)^2+0.52076\times10^{-9}(T-T_0)^3\,,
\end{align*}
where $\epsilon_{\mathrm{T}}$ is dimensionless, $T$ in Kelvin, $T_{0}=300\,$K and with $0.12\%$ mean absolute relative deviations from the respective data. This total hemispherical emissivity dataset correlates reasonably well with published total normal emissivity data\,\cite{emissivit2}. It is important to note that the empirical $\epsilon_{\mathrm{T}}$ expression exhibits very large deviations from available theoretical $\epsilon_{\mathrm{T}}$ calculations\,\cite{emissivit7}. This strong discrepancy is expected, since the classical electromagnetic theory calculations reported in Ref.\cite{emissivit7} are based on the oversimplifying assumption of a single Drude model for the Be dielectric function and on the ad-hoc assumption that only the Drude relaxation time is temperature-dependent through its connection with the resistivity\,\cite{emissivit8,emissivit9}.

For the liquid phase, no total hemispherical emissivity measurements are available in the literature for Be. Given the small $\epsilon_{\mathrm{T}}$ discontinuity at the melting point and weak $\epsilon_{\mathrm{T}}$ temperature dependency of liquid metals\,\cite{viscosity4}, it is assumed that the total hemispherical emissivity remains constant at its melting temperature value from the side of the crystal. This leads to the extrapolation
\begin{align*}
&\epsilon_{\mathrm{T}}(T)=0.811\,.
\end{align*}

\begin{figure}[!b]
         \centering
         \includegraphics[width=3.3in]{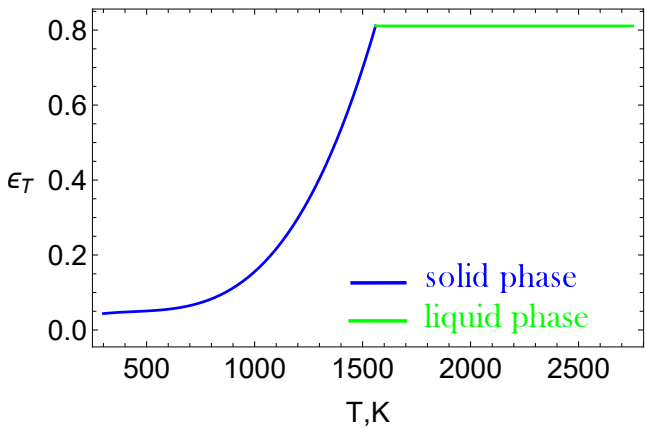}
\caption{Recommended description of the Be \emph{total hemispherical emissivity} in the temperature range $300-2750\,$K. Note that there is no $\epsilon_{\mathrm{T}}$ discontinuity at the liquid-solid transition by construction.}\label{figure_emissivity_recommended}
\end{figure}

Summing up, our recommended description of the beryllium total hemispherical emissivity consists of a cubic polynomial around the room temperature for the solid phase (alpha and beta) and an extrapolated constant value that guarantees phase transition continuity for the liquid phase. This ad hoc extrapolation does not constitute an important source of error, since radiative cooling is always overpowered above the melting point by the exponential nature of vaporization cooling and of thermionic cooling\,\cite{kthmemos03}. Our recommended description is plotted in Figure \ref{figure_emissivity_recommended}.

\subsection{Absolute thermoelectric power}\label{thermoelectric}

\noindent The absolute thermoelectric power emerges in the thermoelectric magnetohydrodynamic description of metallic melts\,\cite{kthmemos01,kthmemos03}. In particular, the Seebeck effect component of Ohm's law is proportional to the absolute thermoelectric power and the Thomson effect component of the heating equation is proportional to the temperature derivative of the absolute thermoelectric power\,\cite{TEMHDrevie}. It is worth pointing out that the mismatch between the absolute thermoelectric powers of a condensed matter PFC (solid or liquid) and its surrounding vapor has been recently conjectured to generate a thermoelectric current that could drive metallic melt motion in extreme tokamak scenarios\,\cite{Seebeckco0}.

For the solid phase at elevated temperatures, only an unpublished polycrystalline Be dataset from Lillie (1955) is quoted in different handbooks\,\cite{resistivi2,emissivit4,emissivit5} that refers to the relative thermoelectric power against platinum within the temperature range $600-1050\,$K. The experimental curve is available in Refs.\cite{emissivit4,emissivit5}, it has been digitized with the aid of software and fitted with a linear function around the room temperature that reads as
\begin{align*}
&S_{\mathrm{rel,Be-Pt}}(T)=1.4275+21.772\times10^{-3}(T-T_0)\,,
\end{align*}
with $S_{\mathrm{rel,Be-Pt}}$ in $\mu$V/K, $T$ in Kelvin, $T_{0}=300\,$K and $0.14\%$ mean absolute relative deviations from the respective data. The absolute thermoelectric power of platinum is described by a ninth-order polynomial\,\cite{Seebeckco1} that is based on a reliable measurement set in the range $273-1600\,$K\,\cite{Seebeckco2} and reads as
\begin{align*}
S_{\mathrm{abs,Pt}}(T)&=22.1561-0.23328T+9.7007\times10^{-4}T^2\\&\,\,\,\,\,\,\,-2.68107\times10^{-6}T^3+4.80917\times10^{-9}T^4\\&\,\,\,\,\,\,\,-5.60664\times10^{-12}T^5+4.18456\times10^{-15}T^6\\&\,\,\,\,\,\,\,-1.92118\times10^{-18}T^7+4.93085\times10^{-22}T^8\\&\,\,\,\,\,\,\,-5.40695\times10^{-26}T^9\,,
\end{align*}
in the same units. In order to obtain the absolute thermoelectric power of beryllium, the two empirical expressions were subtracted, the result was digitized with steps of $50\,$K and the resulting dataset was fitted into a cubic polynomial around the room temperature. Eventually, the empirical expression reads as
\begin{align*}
&S_{\mathrm{abs,Be}}(T)=6.75364+43.0937\times10^{-3}(T-T_0)\\&\,\,\,-8.21233\times10^{-6}(T-T_0)^2+3.17939\times10^{-9}(T-T_0)^3\,,
\end{align*}
with $S_{\mathrm{abs,Be}}$ measured in $\mu$V/K, $T$ in Kelvin, $T_{0}=300\,$K. The absolute thermoelectric power values seem to be high, but are similar to those of other alkanine earth metals\,\cite{Seebeckco3}.

For the liquid phase, no absolute thermoelectric power measurements are available in the literature for Be. Given the typically small $S$ discontinuity at the melting point (positive or negative jumps) and the typically weak $S$ temperature dependency (positive or negative slopes) of liquid metals\,\cite{Seebeckco4,Seebeckco5,Seebeckco6,Seebeckco7}, it is assumed that the absolute thermoelectric power remains constant at its melting temperature value from the side of the crystal. This leads to the extrapolation
\begin{align*}
&S_{\mathrm{abs,Be}}(T)=54.4\,\mu\mathrm{V/K}\,.
\end{align*}

\begin{figure}[!t]
         \centering
         \includegraphics[width=3.3in]{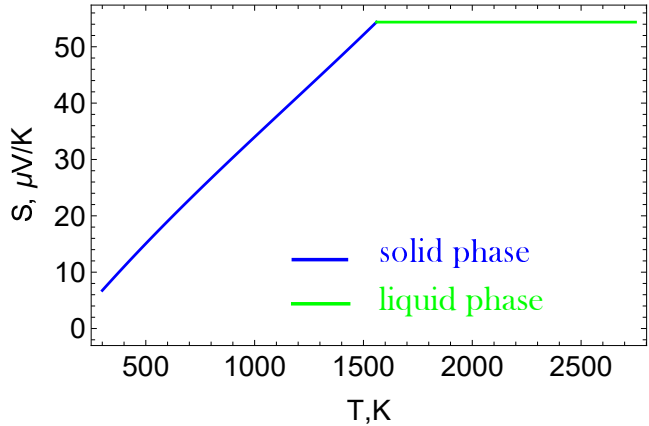}
\caption{Recommended description of the Be \emph{absolute thermoelectric power} in the temperature range $300-2750\,$K. Note that there is no $S$ discontinuity at the liquid-solid transition by construction.}\label{figure_seebeck_recommended}
\end{figure}

Summing up, our recommended description of the beryllium absolute thermoelectric power consists of a cubic polynomial around the room temperature for the solid phase (alpha and beta) and an extrapolated constant value that guarantees phase transition continuity for the liquid phase. Our recommended description is plotted in Figure \ref{figure_seebeck_recommended}.

\section{Discussion and summary}\label{outro}

\subsection{Impurity levels and production methods}\label{miscellaneous}

\noindent In contrast to mechanical properties such as the ultimate tensile strength and the total elongation as well as thermo-mechanical properties such as the thermal shock and thermal fatigue resistance, the thermophysical properties of interest are not very sensitive to the impurity levels, grain size and methods of production. Regardless of the metal, it is well-known that impurity traces at the $\ll1\%$ level are insignificant for thermophysical properties provided that the impurities are not magnetic and the temperatures of interest are above $20\,$K\,\cite{impurity01}.

Since our interest lies on fusion applications, the recommended analytical expressions should be applicable to nuclear Be grades and particularly to ITER reference Be grades (S-65, CN-G01, TGP-56FW) which are characterized by at least $99\%$ minimum purity\,\cite{impurity02}. Consequently, all recommended expressions were based on experimental data for well-annealed or hot-pressed high-purity ($>99\%$) isotropic beryllium. Provided that the purity is high, small impurity variations can be expected to lead to insignificant thermophysical variations that are lower than the measurement uncertainties of most properties\,\cite{impurity03}. Thus, our recommended expressions should also be applicable to some standard grades (such as S-200 of $>98.5\%$). It should be emphasized through that some thermophysical properties such as the thermal conductivity have a strong dependence on porosity\,\cite{impurity04}, thus the respective recommended expressions would need to be modified in scenarios that involve Be foam or plasma sprayed Be\,\cite{impurity05}. It is evident that the same should apply to Be deposition layers as well as to Be flakes that originate from their delamination.

The rather weak impurity dependence of the thermophysical properties can be better understood on a microscopic basis. In metals, electron transport is mainly inhibited by interactions with phonons and interactions with atomic impurities, crystal boundaries and lattice imperfections. These types of collisional contributions are weakly coupled, as expressed by Matthiessen's rule for the electrical resistivity. With the exception of extreme cases where the impurity concentration or dislocation density is very high, electron-phonon collisions dominate already from the room temperature\,\cite{impurity06,impurity07}. Hence, within the temperature range of interest, the \emph{electrical resistivity}, the \emph{thermal conductivity} and the \emph{absolute thermoelectric power} should be weakly dependent on impurities and defects. In addition, unless the temperatures are in the neighborhood of absolute zero, the internal energy of metals is dominated by the vibrational energy of the lattice (see the Debye phonon model) with the valence electron contribution being much smaller (see the free electron theory of Sommerfeld). It is empirically known that the thermodynamically connected \emph{specific heat capacity} and \emph{latent heats} depend primarily on the host lattice and not on trace impurities and lattice imperfections except at the lowest temperatures\,\cite{impurity08}. The same applies to the volume expansion and \emph{mass density}. At this point, it should be emphasized that the solid Be thermal expansion should be anisotropic due to the hcp lattice. However, for isotropic grades, such as the nuclear grades of interest, the differences in thermal expansion along different directions are of the order of few percent\,\cite{impurity09}. Furthermore, the low frequency permittivity of metals (relevant for thermal radiation) is dominated by the Drude contribution of the valence electrons which features the collisional frequency with phonons. Therefore, also the \emph{total hemispherical emissivity} should be weakly dependent on impurities and defects. Finally, the \emph{dynamic viscosity}, the \emph{surface tension}, the \emph{work function} and the \emph{vapor pressure} are either defined only in the liquid state (former two) or important only in the liquid state (latter two), where the production methods and even the impurity levels are much less relevant.

\subsection{Complications in fusion environments}\label{complications}

\noindent In the harsh ITER edge conditions, the Be first wall will be constantly bombarded by hydrogen isotopes and helium (either ions or energetic charge-exchange neutrals), part of which will be absorbed and then get trapped in lattice defects; a phenomenon known as retention\,\cite{retention0,retention1}. Large concentrations of such plasma-induced defects could have an effect on the Be thermophysical properties. Moreover, oxygen will be an intrinsic impurity in ITER\,\cite{researchpl} as in any fusion device and Be is a good getter given its high reactivity with oxygen, which imply that substantial amounts of beryllium oxide (BeO) could be produced. In addition, physical sputtering or chemically assisted physical sputtering of the Be first wall, subsequent ionization and transport by scrape-off layer flows is anticipated to lead to Be deposition on the W divertor, where Be-W intermetallic compounds (Be$_{22}$W, Be$_{12}$W, Be$_{2}$W) of variable thickness could be formed.

It has been well-documented through experiments and simulations that hydrogen retention, helium implantation and plasma-irradiation defects (dislocations, voids and gas-filled bubbles) can lead to a drastic reduction of the tungsten thermal conductivity\,\cite{retention2,retention3,retention4,retention5}. A similar reduction in the thermal conductivity has been experimentally deduced for Be\,\cite{retention6,retention7}. This alone should have an impact on the Be PFC power-handling capabilities leading to unforeseen melting events. More important, plasma retention and irradiation should also effect all other Be thermophysical properties. Nevertheless, trapped gas desorption and nano-structure annealing can be expected to strongly limit such retention and irradiation effects at high temperatures\,\cite{retention8,retention9}. In particular, within the Be liquid phase, dislocations should essentially vanish and even hydrogen or helium that is retained in energetically deep irradiation induced defects should become mobile; diffusing either towards the surface to desorb or into the bulk thus lowering the local impurity density. Further R\&D is necessary to quantify the plasma retention / irradiation effect on different Be thermophysical properties, its dependence on the temperature and the plasma conditions as well as the local extent of thermophysical property degradation. It is evident that it is impossible to cover the entire parameter space for all thermophysical properties. Thus, general trends and the most-affected properties should be at least determined, so that realistic worst-case-scenarios are formulated for different plasma-surface interaction codes. The same considerations apply to neutron-irradiation induced defects that can also lead to a degradation of the Be thermophysical properties, something that has not been systematically studied for Be\,\cite{retention1,neutronic1,neutronic2}. It is possible though that neutron irradiation does not influence most thermophysical properties at the relatively low fluence anticipated in ITER\,\cite{retention0,neutronic1}.

Experience with JET tokamak operation with an ITER-Like-Wall (ILW) has confirmed that BeO formation is mainly promoted in molten Be areas, where surface Be diffusion replenishes the reaction sites available for oxygen\,\cite{beryoxide1}. In JET, the most melting prone Be areas have been identified to be the upper dump plates, where heating to elevated temperatures is caused mainly by unmitigated plasma disruptions\,\cite{beryoxide2}. However, the thickness of these BeO near-surface layers has been measured to be of the order of microns\,\cite{beryoxide1}, which is much thinner than the typical Be melt layer thickness of several hundreds of microns\,\cite{kthmemos02,beryoxide2} and also thinner than the estimated Be ejected droplet size of the order of hundred microns\,\cite{beryoxide2,beryoxide3}. These indicate that there is no real benefit in attempting to consider the presence of BeO in heating, macroscopic melt motion and dust generation studies in JET, especially given the fact that the temperature dependence of most thermophysical properties of BeO has not been investigated yet. On the other hand, the presence of BeO might be relevant in dust transport studies given the typical Be size distributions collected in JET\,\cite{beryoxide4,beryoxide5,beryoxide6}. In general, it cannot be excluded that the more intense plasma-surface interactions in ITER lead to the formation of thicker BeO layers. In that case, the presence of BeO would still be inconsequential in heating, macroscopic melt motion and dust generation simulations but might be relevant to arcing and to dust transport simulations. For such studies, the change in the role of thermionic emission should be rather critical. In particular, thermionic emission is negligible for Be owing to its high work function of $4.98\,$eV and its low melting point of $1560\,$K, but thermionic emission is likely to be very important for BeO owing to its low work function of $3.60\,$eV and its high melting point of $2780\,$K.

The JET-ILW succesfull operation in a mixed Be/W environment has confirmed the formation of deposited Be layers and revealed the formation of Be-W intermetallic compounds in the W divertor\,\cite{intermeta1,intermeta2}. In addition, the post-mortem analysis of a Be limiter tile has revealed that different Be-W intermetallic compounds are formed in the deposition zone, but that the erosion zone contains pure Be\,\cite{intermeta3}. Finally, Be-W intermetallic compounds of variable stoichiometry might also be formed below the contact areas of Be dust adhered to hot W divertor areas\,\cite{intermeta4}. In all the above cases, the thickness of the Be-W intermetallic compounds lies at the sub-micron range, which indicates that the largely unknown Be$_{22}$W, Be$_{12}$W or Be$_{2}$W thermophysical properties do not need to be considered.

Finally, it should be noted that the total hemispherical emissivity, at least for low temperatures, has a strong dependence on the surface roughness, the deuterium content and the oxygen concentration\,\cite{thermogra1,thermogra2}. Thus, an \emph{in-situ} assessment of the Be emissivity is preferable, especially for IR thermography studies\,\cite{thermogra3,thermogra4}, and our recommended total hemispherical emissivity expression should only be employed when such an option is not available.

\subsection{Comparison with the ITER Materials Properties Handbook}\label{handbook}

\noindent The ITER Materials Properties Handbook is mainly focused on mechanical properties. Unfortunately, it contains merely an elementary presentation of basic thermophysical properties of solid Be and provides even less information on the thermophysical properties of liquid Be\,\cite{ITERhandb1,ITERhandb2}. In particular, liquid Be recommendations are only available for the specific heat capacity and solid Be recommendations are typically based on outdated rather than state-of-the-art measurements. For instance, the majority of the presented Be datasets are adopted from the TPRC Data Series that was edited by Touloukian in the 70s. A detailed comparison between the ITER recommendations and the present recommendations can be found in what follows.
\begin{itemize}
\item \emph{Phase transition temperatures and latent heats}. The $1560\,$K and $2744\,$K ITER recommendations for the melting and normal boiling point are very accurate. There is no discussion on the polymorphic transition. No ITER recommendations are available for the enthalpies of transformation, fusion and vaporization.
\item \emph{Specific isobaric heat capacity}. For the solid phase, the recommended ITER analytical description exhibits small but observable deviations from our recommendation in the entire temperature range. The largest relative deviations are present in the vicinity of $400\,$K ($4\%$) and near the melting point ($2\%$). For the liquid phase, the ITER recommendation is nearly indistinguishable from our recommendation, with $0.6\%$ maximum relative deviations at the normal boiling point.
\item \emph{Electrical resistivity}. For the solid phase, the recommended ITER analytical description exhibits deviations from our recommendation in the entire temperature range. The relative deviations monotonically decrease from $25\%$ at room temperature up to $8\%$ at the Be melting point. For the liquid phase, there is no ITER recommendation.
\item \emph{Thermal conductivity}. For the solid phase, the recommended ITER analytical description is very close to our recommendation up to $1000\,$K (deviations less than $2\%$). However, for higher temperatures, the relative deviations increase up to $20\%$. Prior to the Be melting point, the ITER expression even has an unphysical maximum that is followed by an equally unphysical monotonic increase with the temperature. For the liquid phase, there is no ITER recommendation.
\item \emph{Mass density}. For the solid phase, the recommended ITER analytical description is very close to our recommendation in the entire temperature range (deviations less than $1.2\%$), in spite of the use of the erroneous room temperature mass density of $\rho_{\mathrm{m0}}=1.83\,$g/cm${^3}$. Correction to $\rho_{\mathrm{m0}}=1.85\,$g/cm${^3}$ leads to nearly indistinguishable results up to $1100\,$K and less than $0.5\%$ deviations up to the Be melting point. For the liquid phase, there is no ITER recommendation.
\item \emph{Surface tension}. There is no ITER recommendation.
\item \emph{Dynamic viscosity}. There is no ITER recommendation.
\item \emph{Vapor pressure}. The recommended ITER analytical description begins to exhibit deviations from our recommended Antoine correlation above $1800\,$K. These deviations monotonically increase with the temperature and even reach a factor of four in the vicinity of the normal Be boiling point. Essentially, the ITER recommendation significantly overestimates the vapor pressure for all those temperatures for which vaporization is important.
\item \emph{Work function}. There is no ITER recommendation.
\item \emph{Total hemispherical emissivity}. For the solid phase, some datasets are presented but are not accompanied by any recommendation. It is worth noting that at least one dataset refers to the total normal emissivity and not the total hemispherical emissivity. For the liquid phase, there is no ITER recommendation.
\item \emph{Thermoelectric power}. There is no ITER recommendation.
\end{itemize}

\subsection{Status of the literature}\label{summary}

\begin{table*}[!t]
  \centering
  \caption{Analytical description of the thermophysical properties of solid polycrystalline beryllium from the vicinity of the room temperature up to the melting temperature and the thermophysical properties of liquid beryllium from the melting temperature up to the normal boiling temperature. The physical properties of interest are the enthalpy of alpha to beta transition ($\Delta{h}_{\mathrm{t}}$), the enthalpy of fusion ($\Delta{h}_{\mathrm{f}}$), the enthalpy of vaporization ($\Delta{h}_{\mathrm{v}}$), the specific isobaric heat capacity ($c_{\mathrm{p}}$), the electrical resistivity ($\rho_{\mathrm{e}}$), the thermal conductivity ($k$), the mass density ($\rho_{\mathrm{m}}$), the surface tension ($\sigma$), the dynamic viscosity ($\mu$), the vapor pressure ($p_{\mathrm{v}}$), the work function ($W_{\mathrm{f}}$), the total hemispherical emissivity ($\epsilon_{\mathrm{T}}$) and the absolute thermoelectric power ($S$). In the empirical expressions below: $T_0=300\,$K, $T_{\mathrm{t}}=1543\,$K, $T_{\mathrm{m}}=1560\,$K, $T_{\mathrm{b,n}}=2750\,$K and $T_{\mathrm{c}}=8080\,$K.}\label{tablethermophysical}
  \begin{tabular}{c c c c c} \hline
Units             & \,\,Recommended description\,\,                                                                                                                                        & Ref.              \\ \hline
kJ/mol            & $\Delta{h}_{\mathrm{t}}=6.855$\qquad\qquad\qquad\qquad\qquad\qquad\qquad\qquad\qquad\qquad\qquad\qquad\qquad\qquad\qquad\qquad\qquad\qquad\qquad\quad\,\,              & \cite{Thermodyn4} \\
kJ/mol            & $\Delta{h}_{\mathrm{f}}=7.959$\qquad\qquad\qquad\qquad\qquad\qquad\qquad\qquad\qquad\qquad\qquad\qquad\qquad\qquad\qquad\qquad\qquad\qquad\qquad\quad\,\,              & \cite{Thermodyn4} \\
kJ/mol            & $\Delta{h}_{\mathrm{v}}(T>T_0)=324[(T_{\mathrm{c}}-T)(T_{\mathrm{c}}-T_0)]^{0.28}$\qquad\qquad\qquad\qquad\qquad\qquad\qquad\qquad\qquad\qquad\qquad\qquad\qquad\,\,\, & \cite{Thermodyn8} \\
J/(mol$\cdot$K)   & $c_{\mathrm{p}}(T>T_0)=21.205+5.694\times10^{-3}T+0.962\times10^{-6}\times{T}^2-0.5874\times10^6/T^2$\qquad\qquad\qquad\qquad\qquad\,\,\,                              & \cite{heatcapac1} \\
J/(mol$\cdot$K)   & $c_{\mathrm{p}}(T>T_{\mathrm{t}})=30$\qquad\qquad\qquad\qquad\qquad\qquad\qquad\qquad\qquad\qquad\qquad\qquad\qquad\qquad\qquad\qquad\qquad\qquad\quad\,\,             & \cite{Thermodyn4} \\
J/(mol$\cdot$K)   & $c_{\mathrm{p}}(T>T_{\mathrm{m}})=25.4345+2.150\times10^{-3}T\qquad\qquad\qquad\qquad\qquad\qquad\qquad\qquad\qquad\qquad\qquad\qquad\qquad\quad\,$                    & \cite{Thermodyn2} \\
$\mu\Omega-$cm    & $\rho_{\mathrm{e}}(T>T_0)=3.71002+30.4119\times10^{-3}(T-T_0)+2.7851\times10^{-6}(T-T_0)^2+3.25184\times10^{-9}(T-T_0)^3$                                              & \cite{resistivi1} \\
$\mu\Omega-$cm    & $\rho_{\mathrm{e}}(T>T_{\mathrm{m}})=45\qquad\qquad\qquad\qquad\qquad\qquad\qquad\qquad\qquad\qquad\qquad\qquad\qquad\qquad\qquad\qquad\qquad\qquad\,\,\,\,\,\,$       & extrap            \\
W/(m$\cdot$K)     & $k(T>T_0)=148.8912-76.3780\times10^{-3}T+12.0174\times10^{-6}T^2+6.5407\times10^6/T^2\qquad\qquad\qquad\qquad\,\,\,\,\,\,\,$                                           & \cite{conductiv1} \\
W/(m$\cdot$K)     & $k(T>T_{\mathrm{m}})=84.59+54.22\times10^{-3}(T-T_{\mathrm{m}})\qquad\qquad\qquad\qquad\qquad\qquad\qquad\qquad\qquad\qquad\qquad\qquad\,\,\,\,\,\,\,\,$               & extrap            \\
g/cm${^3}$        & $\rho_{\mathrm{m}}(T>T_0)=1.850-6.8648\times10^{-5}(T-T_0)-4.1660\times10^{-8}(T-T_0)^2+1.1354\times10^{-11}(T-T_0)^3\quad\,$                                          & \cite{massdensi1} \\
g/cm${^3}$        & $\rho_{\mathrm{m}}(T>T_{\mathrm{m}})=1.690-0.116\times10^{-3}(T-T_{\mathrm{m}})\qquad\qquad\qquad\qquad\qquad\qquad\qquad\qquad\qquad\qquad\qquad\qquad\,\,\,$         & \cite{massdensi2} \\
N/m               & $\sigma(T>T_{\mathrm{m}})=1.143-0.20\times10^{-3}(T-T_{\mathrm{m}})\qquad\qquad\qquad\qquad\qquad\qquad\qquad\qquad\qquad\qquad\qquad\qquad\quad\,\,\,\,$              & \cite{Thermodyn7} \\
Pa$\cdot$s        & $\mu(T>T_{\mathrm{m}})=0.514\times10^{-3}\exp{\left(4.635T_{\mathrm{m}}/T\right)}\qquad\qquad\qquad\qquad\qquad\qquad\qquad\qquad\qquad\qquad\qquad\qquad\quad\,\,\,$  & \cite{viscosity5} \\
Pa                & $\log{p_{\mathrm{v}}(T>T_0)}=10.2089-13696.6102/(T-124.63)\qquad\qquad\qquad\qquad\qquad\qquad\qquad\qquad\qquad\qquad\quad\,\,\,\,$                                   & \cite{vaporpres3} \\
eV                & $W_{\mathrm{f}}=4.98\qquad\qquad\qquad\qquad\qquad\qquad\qquad\qquad\qquad\qquad\qquad\qquad\qquad\qquad\qquad\qquad\qquad\qquad\qquad\qquad\,\,$                      & \cite{workfunc06} \\
--                & $\epsilon_{\mathrm{T}}(T>T_0)=0.043865+5.728\times10^{-5}(T-T_0)-2.18399\times10^{-7}(T-T_0)^2+5.2076\times10^{-10}(T-T_0)^3$                                          & \cite{emissivit6} \\
--                & $\epsilon_{\mathrm{T}}(T>T_{\mathrm{m}})=0.811\qquad\qquad\qquad\qquad\qquad\qquad\qquad\qquad\qquad\qquad\qquad\qquad\qquad\qquad\qquad\qquad\qquad\qquad$            & extrap            \\
$\mu\mathrm{V/K}$ & $S(T>T_0)=6.75364+43.0937\times10^{-3}(T-T_0)-8.21233\times10^{-6}(T-T_0)^2+3.17939\times10^{-9}(T-T_0)^3$                                                             & \cite{emissivit4} \\
$\mu\mathrm{V/K}$ & $S(T>T_{\mathrm{m}})=54.4\qquad\qquad\qquad\qquad\qquad\qquad\qquad\qquad\qquad\qquad\qquad\qquad\qquad\qquad\qquad\qquad\qquad\qquad\quad$                            & extrap            \\ \hline \hline
\end{tabular}
\end{table*}

\noindent The thermophysical properties analyzed in this work constitute input for numerical codes that simulate the thermal, thermo-electric and magneto-hydrodynamic response of beryllium plasma-facing components, including solid dust grains and molten droplets, to incident heat and plasma particle fluxes. The recommended analytical expressions for the temperature dependence of the thermophysical properties of high purity polycrystalline solid and liquid beryllium have been gathered in Table \ref{tablethermophysical}. The summary table is self-containing, in the sense that the interested code developer can directly copy the empirical expressions of one's interest without being pre-occupied with details (only the primary reference has been provided). It is worth mentioning that the macroscopic melt layer motion code MEMOS-U\,\cite{kthmemos01,kthmemos02,kthmemos03} and the dust transport code MIGRAINe\,\cite{migraine01,migraine02,migraine03} have already been updated following the present recommendations.

The status of the experimental datasets for the thermophysical properties of solid beryllium is satisfying. Multiple independent measurements are available for the specific enthalpy (leading to the accurate determination of the latent heats and of the specific isobaric heat capacity), the electrical resistivity, the thermal conductivity, the mass density, the vapor pressure as well as the work function. Unfortunately, single extended-in-temperature measurements are available for the total hemispherical emissivity and the absolute thermoelectric power. Independent datasets, that are preferably obtained with different techniques, are certainly desirable for the latter two physical quantities.

On the other hand, the status of the experimental data-sets for the thermophysical properties of liquid beryllium is substandard. Extrapolations had to be carried out for the electrical resistivity, the thermal conductivity, the total hemispherical emissivity and the absolute thermoelectric power. Moreover, from the two extended-in-temperature measurements that are available for the specific enthalpy and the mass density, only one dataset was judged to be reliable. Finally, single extended-in-temperature measurements are available for the surface tension and for the dynamic viscosity. In general, new measurements based on state-of-the-art techniques (e.g levitation calorimetry) are rather imperative for most thermophysical quantities. The extrapolations attempted herein are safe close to the melting point, but should lead to progressively larger errors as the temperature further increases towards the normal boiling point.

\section*{Acknowledgments}

\noindent The author acknowledges the financial support of the Swedish Research Council under Grant No 2021-05649. This work has been carried out within the framework of the EUROfusion Consortium and funded by the European Union via the Euratom Research and Training Programme (Grant Agreement No 101052200 — EUROfusion). Views and opinions expressed are however those of the author only and do not necessarily reflect those of the European Union or European Commission. Neither the European Union nor the European Commission can be held responsible for them.

\end{document}